\documentclass[twocolumn, tighten]{aastex63}

\usepackage{color, amsmath, natbib}
\hypersetup{breaklinks=true, colorlinks=true, urlcolor={blue}, citecolor={blue}, linkcolor={blue}}

\usepackage{etoolbox}
\shortauthors{Hu et al.}
\accepted{Dec.~29, 2020}

\makeatletter
\patchcmd{\NAT@citex}
  {\@citea\NAT@hyper@{%
     \NAT@nmfmt{\NAT@nm}%
     \hyper@natlinkbreak{\NAT@aysep\NAT@spacechar}{\@citeb\@extra@b@citeb}%
     \NAT@date}}
  {\@citea\NAT@nmfmt{\NAT@nm}%
   \NAT@aysep\NAT@spacechar\NAT@hyper@{\NAT@date}}{}{}

\patchcmd{\NAT@citex}
  {\@citea\NAT@hyper@{%
     \NAT@nmfmt{\NAT@nm}%
     \hyper@natlinkbreak{\NAT@spacechar\NAT@@open\if*#1*\else#1\NAT@spacechar\fi}%
       {\@citeb\@extra@b@citeb}%
     \NAT@date}}
  {\@citea\NAT@nmfmt{\NAT@nm}%
   \NAT@spacechar\NAT@@open\if*#1*\else#1\NAT@spacechar\fi\NAT@hyper@{\NAT@date}}
  {}{}

\makeatother

\newcommand{\LD}[1]{{\color{black} #1}}

\newcommand{\chandra}{\emph{Chandra}}

\newcommand{\xmm}{\emph{XMM-Newton}}

\newcommand{\lumcgs}{erg~s$^{-1}$}

\begin{document}
\title{Possible Periodic Dips in the Pulsating Ultraluminous X-ray Source M51 ULX-7}
\author[0000-0001-8551-2002]{Chin-Ping Hu}
\altaffiliation{JSPS International Research Fellow}
\affiliation{Department of Physics, National Changhua University of Education, Changhua, 50007, Taiwan}
\affiliation{Department of Astronomy, Kyoto University, Kitashirakawa-Oiwake-cho, Sakyo-ku, Kyoto 606-8502, Japan}
\author[0000-0001-7821-6715]{Yoshihiro Ueda}
\affiliation{Department of Astronomy, Kyoto University, Kitashirakawa-Oiwake-cho, Sakyo-ku, Kyoto 606-8502, Japan}
\author[0000-0003-1244-3100]{Teruaki Enoto}
\affiliation{Extreme Natural Phenomena RIKEN Hakubi Research Team, Cluster for Pioneering Research, RIKEN, 2-1 Hirosawa, Wako, Saitama 351-0198, Japan}

\correspondingauthor{C.-P. Hu}
\email{cphu0821@cc.ncue.edu.tw}

\begin{abstract}
We report the discovery of possible periodic X-ray dips in a pulsating ultraluminous X-ray source, M51 ULX-7, with the archival \chandra\ observations. With $\sim$20 days of monitoring in the superorbital descending state, we discovered three dips with separations of $\sim$2 and $\sim$8 days via the Bayesian block technique. A phase-dispersion minimization and a $\chi^2$ test suggest that the dip is likely recurrent with a period of $\sim2$ days, consistent with the orbital period of M51 ULX-7. We interpret the dip as an obscuring of the emission from the pulsar by the vertical structure on the stream-disk interaction region or the atmosphere of the companion star. Both interpretations suggest the viewing angle to be $\sim60$ degrees. Given that the magnetic field of M51 ULX-7 is moderately high, $B\sim10^{13}$ G, a low geometric beaming with $b\lesssim1/2$ is sufficient to explain the observed flux and the presence of dips. Obscuration of the stellar wind remains an alternative possible origin and further monitoring of the dips will be required. 


\end{abstract}


\section{Introduction}\label{introduction}
Ultraluminous X-ray sources (ULXs) are extragalactic off-nuclear X-ray point sources with luminosities exceeding the Eddington limit of a stellar-mass black hole (BH) \citep{FengS2011, KaaretFR2017}. They are the most extreme populations of compact objects that break several theoretical limits. The most remarkable one is the Eddington limit, which can be calculated as $L_{\rm{Edd}}=4\pi GMm_{\rm{p}}/\sigma_{\rm{T}}\approx1.3\times10^{38} \left( M/M_{\rm{\odot}} \right)$ erg s$^{-1}$, where $\sigma_{\rm{T}}$ is the Thomson scattering cross-section, $m_{\rm{p}}$ is the mass of a proton, and $M$ is the mass of the accretor. For spherical accretion, this is the upper limit of the luminosity because the strong radiation pressure prevents mass accretion. Intermediate-mass BHs with masses of $10^2$ -- $10^4$ $M_{\odot}$ could explain the observed high luminosity but only a few ULXs are candidates for intermediate-mass BHs \citep{GodetPK2012, StraubGW2014}. Alternatively, most ULXs are likely stellar-mass compact objects with super-Eddington accretion rates and mild beaming through population studies \citep{KaaretFR2017}.

The discovery of pulsating ULXs (PULXs) is a milestone that challenges current understanding of both the ULXs and the magnetospheric accretion of neutron stars (NSs). Until early 2020, eight PULXs have been discovered \citep{TrudolyubovPC2007, Trudolyubov2008, BachettiHW2014, FuerstWH2016, IsraelBS2017, CarpanoHM2018, SathyaprakashRW2019, RodriguezCastilloIB2020}. The Eddington luminosity of an NS is one order of magnitude lower than that of a 10 $M_{\odot}$ BH. PULXs are believed to be powered by NSs with extremely high mass accretion rates and/or extremely high magnetic fields. It was proposed that an NS with an extremely high magnetic field could explain the observed luminosity because the cross-section can be reduced and the emission can radiate along the direction perpendicular to the accreting flow \citep{BaskoS1975, MushtukovST2015}.  Alternatively, a strong magnetic field may not be necessary if their emission could be more strongly collimated compared to ULXs that harbor BHs \citep{KingL2019}. Most of the currently known PULXs can be well interpreted with the super-critical accretion model in which the emission is highly beamed toward the Earth \citep{KingL2019}. For example, the half-opening angle of the funnel and the inclination angle of NGC 5907 X-1 are estimated as $\lesssim10^{\circ}$ \citep{DauserMW2017}. 

M51 is an interacting galaxy pair consisting of an active face-on spiral galaxy, NGC 5194, and a dwarf galaxy, NGC 5195. This system contains hundreds of X-ray sources including at least nine ULXs with $L_X>10^{39}$ erg s$^{-1}$ and 13 point X-ray sources with $10^{38}$ erg s$^{-1} < L_X < 10^{39}$ erg s$^{-1}$ \citep{TerashimaW2004}. M51 ULX-7 (also known as CXOU J133001.0+471344 and RX J133001+47137, hereafter ULX-7) is a highly variable ULX located at a spiral arm toward the northwest from the center of NGC 5194. ULX-7 was recently found to be a PULX with a spin period of 2.8 s and an orbital period of $\sim$2 days. The inclination angle is unconstrained due to the lack of X-ray orbital modulations in \xmm\ data \citep{RodriguezCastilloIB2020}. A superorbital modulation with a period of $\approx38$ days can be observed with the monitoring data taken by the \emph{Neil Gehrels Swift} observatory although the mechanism remains unknown \citep{BrightmanEF2020, VasilopoulosLK2020}. The measured spin-up rate ($\dot{\nu}=3.1(8)\times10^{-11}$ Hz s$^{-1}$) suggests that the NS in M51 ULX-7 has a moderately strong magnetic field with $B = 10^{12}$--$10^{13}$ G assuming that the superorbital modulation is caused by changes in the mass accretion rate \citep{RodriguezCastilloIB2020}, or $B=(3$--$4)\times10^{13}$ G if the superorbital modulation is caused by the free precession of the NS \citep{VasilopoulosLK2020}.

\section{Observations and Data Reduction}\label{sec:observation}
\subsection{\chandra}
M51 was observed with the Advanced CCD Imaging Spectrometer (ACIS) of \chandra\ 16 times between 2000 and 2018. Except for ObsID 19522 which was observed with ACIS-I chips, they were observed with ACIS-S chips. Seven of them (ObsIDs: C1: 13813, C2: 13812, C3: 15496, C4:13814, C5: 13815, C6: 13816, and C7: 15553) were intensively carried out between 2012 September and 2012 October (see Table \ref{tab:observation_log}). These data sets allow us to investigate variability with time scales from a few hours to a few days. We reprocess all of the data sets using the pipeline \texttt{chandra\_repro} in the Chandra Interactive Analysis of Observations (CIAO) version 4.9 with the calibration database (CALDB) version 4.7.3 \citep{FruscioneMA2006}. 

We extract source events from a circular aperture of 3\arcsec\ radius in which 95\% of the source photons are encircled. We use \texttt{dmextract} to extract the light curve of ULX-7 in 0.5 -- 7 keV with 7200 s time resolution. The background is extracted from a nearby source-free region and subtracted from the source light curve. The time intervals with background flares are removed by using the \texttt{lc\_clean} command. We find that the count rate sometimes falls to zero or rises to extremely large values at the beginning or end of an observation. Therefore, we remove the boundary points of each observation. 

\subsection{\xmm}
\xmm\ has carried out 10 observations targeting M51 before 2019, while four of them (ObsIDs X1: 0824450901, X2: 0830191401, X3: 0830191501, and X4: 0830191601) with long exposure times were made in 2018 May and June. All the EPIC cameras were operated in the full-frame mode in X1, while the MOS camera switched to small window mode in the remaining three observations to resolve the pulsation signal of ULX-7. We \LD{downloaded} the data from the \emph{XMM-Newton} Science Archive. The pipeline procedures \texttt{epproc} and \texttt{emproc} tasks in the XMM-Newton Science Analysis Software (SAS version 18.0.0) are used to reprocess the PN and MOS data with the latest calibration files, respectively. We do not further filter out any time intervals since no significant background flaring was found in these observations.

We merge the events from all EPIC cameras to increase the signal-to-noise ratio and extract \emph{XMM-Newton} light curves of ULX-7 using the \texttt{epiclccorr} command. The source events are selected from a 15\arcsec\ circle \LD{centered} on ULX-7. This criterion only encircles $\lesssim70$ \% of source energy, \LD{but can help to avoid} contamination from a nearby source, CXOU J133004.3+471321, which is only 40\arcsec\ away from ULX-7. The background is extracted from a source-free region although contamination from diffuse emission from the spiral arm of M51a remains possible. The time bin size of the light curve is set to 3600 s.

\begin{table*}
    \caption{Data sets used in this work. }
    \label{tab:observation_log}
    \footnotesize
    \begin{tabular}{ccccccc}
        \hline
        \hline
        Epoch & Observatory & ObsID & Instrument & Mode &  Start Date & Exposure (ks) \\
        \hline
           C1 & \emph{Chandra} & 13813 & ACIS-S & TE  & 2012-09-09 17:47:30 & 179.2\\
           C2 & \emph{Chandra}    & 13812       & ACIS-S & TE & 2012-09-12 18:23:50 & 157.5 \\
           C3 & \emph{Chandra}    & 15496       & ACIS-S & TE & 2012-09-19 09:20:34 & 41.0  \\
           C4 & \emph{Chandra}    & 13814       & ACIS-S & TE & 2012-09-20 07:21:42 & 189.9 \\
           C5 & \emph{Chandra}    & 13815       & ACIS-S & TE & 2012-09-23 08:12:08 & 67.2  \\
           C6 & \emph{Chandra}    & 13816       & ACIS-S & TE & 2012-09-26 05:11:40 & 73.1  \\
           C7 & \emph{Chandra}    & 15553       & ACIS-S & TE & 2012-10-10 00:43:36 & 37.6  \\
           X1 & \emph{XMM-Newton} & 0824450901  & PN     & FF & 2018-05-13 22:02:37 & 74.8  \\
           &            &             & MOS1   & FF & 2018-05-13 21:36:55 & 76.6  \\
           &            &             & MOS2   & FF & 2018-05-13 21:37:15 & 76.6  \\
           X2 & \emph{XMM-Newton} & 0830191401  & PN     & FF & 2018-05-25 21:10:48 & 94.8  \\
           &            &             & MOS1   & SW & 2018-05-25 20:45:06 & 96.6  \\
           &            &             & MOS2   & SW & 2018-05-25 20:45:28 & 96.6  \\
           X3 & \emph{XMM-Newton} & 0830191501  & PN     & FF & 2018-06-13 02:22:52 & 59.8  \\
            &            &             & MOS1   & SW & 2018-06-13 01:57:11 & 61.6  \\
           &            &             & MOS2   & SW & 2018-06-13 01:57:33 & 61.6  \\
           X4 & \emph{XMM-Newton} & 0830191601  & PN     & FF & 2018-06-15 02:08:11 & 59.8  \\
           &            &             & MOS1   & SW & 2018-06-15 01:42:29 & 61.6  \\
           &            &             & MOS2   & SW & 2018-06-15 01:42:51 & 61.6 \\
         \hline
\tabularnewline
\end{tabular}
\\ {\footnotesize{The abbreviations and time resolution of the observation mode are described as follows. TE: timed exposure mode of full-frame ACIS-S chip has a time resolution of 3.2 s. FF: full-frame mode with time resolution of 73.4 ms for PN and 2.6 s for MOS. SW: small window mode of the MOS camera with
time resolution of 0.3 s. }}
\end{table*}

\section{Analysis and Results}
\subsection{\chandra\ Light Curves}
\begin{figure}
\centering
\includegraphics[width=0.9\linewidth]{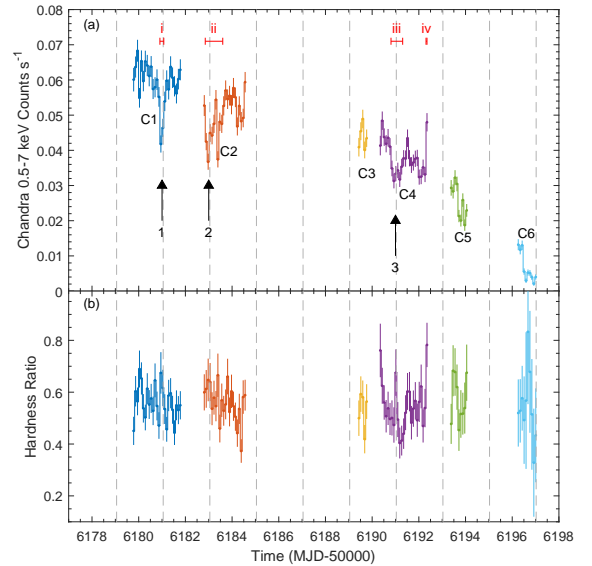} 
\caption{ (a) The \chandra\ 0.5 -- 7 keV light curve of ULX-7 observed in 2012. Six data sets (C1--C6) are plotted. This 20 day light curve was carried out during the descent of the superorbital modulation. Three dips occurred at MJD 56181, 56183, and 56191, indicated by arrows. The gray dashed line denotes the expected dip arrival times by assuming that the dip arrives periodically with a period equal to the orbital period of 1.9969 days. The intervals \rm{i}--\rm{iv} indicate the time intervals during which the Bayesian block analysis identifies significant flux deviation from the long-term trend. (b) Corresponding hardness ratio curve of ULX-7. \label{fig:chandra_lc} }
\end{figure}

Following the standard analysis procedure described in Section \ref{sec:observation}, we create background-subtracted light curves with a bin size of two hours in the 0.5--7 keV band (see Figure \ref{fig:chandra_lc} (a)). The count rate decreases from 0.06 counts s$^{-1}$ to almost zero. This is likely the transition from high to low states in a superorbital cycle. Three dips are seen in observations C1 (dip1), C2 (dip2), and C4 (dip3). Dip 1 and dip 3 are fully observed, while dip 2 is partially seen. 

To test whether the dips are statistically significant, we utilize the Bayesian block\footnote{\url{https://docs.astropy.org/en/stable/api/astropy.stats.bayesian_blocks.html}} technique to search for significant flux variability \citep{ScargleNJ2013}. Data sets C5 and C6 are not included in this analysis because they are likely observed in the superorbital low low sate. The flux variability in other data sets is heavily dominated by the flux decrease in the superorbital cycle. We normalize the count rate of each observation with respect to the superorbital decreasing trend, which is obtained by smoothing the light curve in Figure \ref{fig:chandra_lc}(a), to make the average value to be unity. We also tried to obtain the trend individually by fitting each data set with a linear function and obtaining a consistent result. With a false alarm probability of $p_0=0.003$, we identified four significant flux variations. Three of them (\rm{i}, \rm{ii}, and \rm{iii} in Figure \ref{fig:chandra_lc}) correspond to the visible dips, implying that the flux of these dips significantly deviates from the long-term flux trend. The center time of dip 1 and dip 3 are simply calculated from the center of the blocks as MJD 56180.99 and 56191.06, respectively. The fourth (\rm{iv}) flux variation corresponds to a sudden flux increment in the last time bin of the C4 observation. This could be caused by flares of ULX-7 or unknown instrumental effects, which is beyond the scope of this paper. 

As an alternative method to estimate dip parameters, we fit the count rate during the non-dip state with a straight line and estimate the dip center time ($t_{\rm{c}}$) as the summation of the time of all the bins in the dip with a weighting factor defined as the difference between measured count rate and estimated non-dip count rate \citep{HuCC2008}.  We find that $t_{(\rm{c,dip1})}$=MJD $56180.99\pm0.03$ and $t_{(\rm{c,dip3})}$=MJD $56191.06\pm0.03$, consistent with those directly obtained from the blocks. Dip 2 is not fully observed but the center is likely located around MJD 56183. The time separation between dip 1 and 2 is two days and that between dip 2 and dip 3 is eight days. Given that the orbital period of M51 ULX-7 is two days, we suggest that this dip feature appeared periodically per binary orbit. The widths of the dips are calculated as the dispersion, i.e., the second moment of the dip state, as $W_{\rm{dip1}}=0.07\pm0.03$ days and $W_{\rm{dip3}}=0.16\pm0.02$ days. This hints that the width could be superorbital phase-dependent. 

\begin{figure}
\centering
\includegraphics[width=0.9\linewidth]{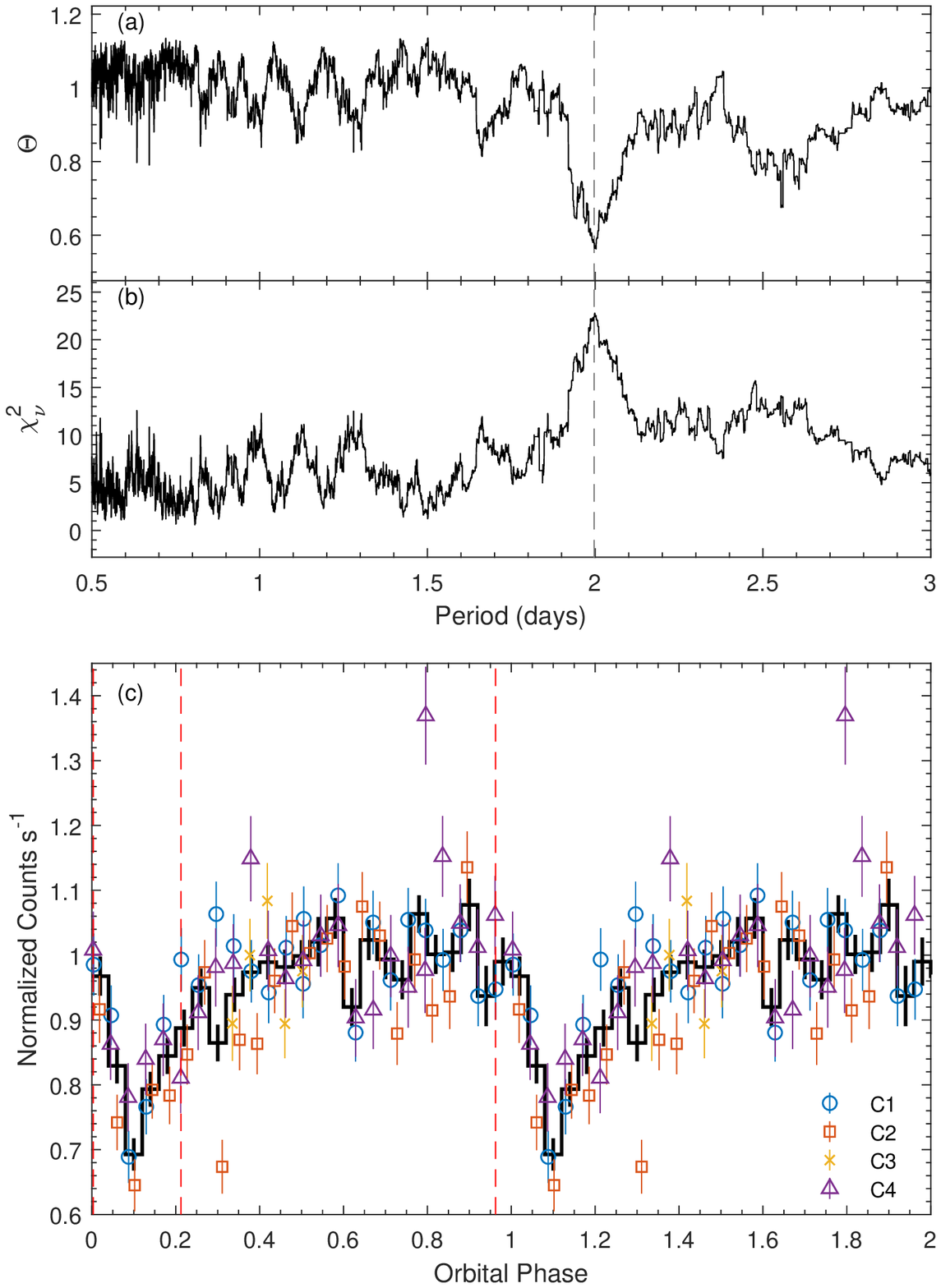} 
\caption{ (a) Phase dispersion minimization and (b) the epoch folding period search results of the normalized \chandra\ light curve of ULX-7. The vertical dashed line is the orbital period derived from the orbital Doppler effect of the pulsar \citep{RodriguezCastilloIB2020}. (c) Normalized \chandra\ light curve folded with the orbital period of ULX-7. The scattered data points represent data points from data sets C1--C4, while the black histogram denotes the mean orbital profile by dividing an orbital cycle into 24 bins. A dip can be seen at orbital phase $\phi_{\rm{orb}}\approx0.1$.  The red dashed lines are edges where the Bayesian block technique identifies a significant change in the count rate.  \label{fig:chandra_fold_lc}}
\end{figure}

To test whether the dip originates from the orbital motion, we searched for periodicity with two non-Fourier-based algorithms, the phase dispersion minimization \citep{Stellingwerf1978} and the epoch folding period search \citep{LeahyDE1983} applied to the detrended \chandra\ light curve of ULX-7. The result is shown in Figure \ref{fig:chandra_fold_lc}. Both algorithms show a signal around the known orbital period of $\sim2$ days \citep{RodriguezCastilloIB2020}.

To test whether the spectral behavior of ULX-7 changes during the dip, we calculate the hardness ratio ($HR$) defined as $HR= (\rm{r}_h-\rm{r}_s)/(\rm{r}_h+\rm{r}_s)$, where r$_h$ is the count rate in the 2--7 keV band and r$_s$ is that in the count rate at 0.5--2 keV (see Figure \ref{fig:chandra_lc} (b)). We find no clear hardening during the dip, similar to a few other dipping systems in the same galaxy \citep{UrquhartS2016, WangSU2018}. 


We further folded the light curve with the binary orbital period to obtain the orbital profile. The phase zero epoch is set to MJD 58285.0084 (2018 June 16), which is the superior conjunction when the pulsar is behind the companion star.  The unbinned light curve and binned orbital profile are plotted in Figure \ref{fig:chandra_fold_lc} (c). A clear orbital dip can be seen with a dip center phase at 0.1. To test whether the dip is significantly determined, we again used the Bayseian block technique to search for significant variability in the unbinned folded light curve (see the colored data points in Figure \ref{fig:chandra_fold_lc}). We searched for optimized blocks by setting the false alarm probability $p_0$ to be 0.003. The result shows that an entire orbital cycle can be divided into two blocks, including a dip at $\phi_{\rm{orb}}<0.21$ and a persistent phase at $\phi_{\rm{orb}}>0.21$


\subsection{\chandra\ Spectral Analysis}
We then investigate X-ray spectral behaviors during the dip states and the non-dip states. We use \texttt{dmgti} to split the event files into the dip and non-dip states, and use \texttt{specextract} to extract the X-ray source and background spectra and corresponding response files and ancillary files. The spectral fitting is achieved using \LD{CIAO's} modeling package \LD{SHERPA}. We fit the 0.3 -- 10 keV spectra using the Cash statistics \citep{Cash1979}. We choose the absorption model \texttt{tbabs} and the corresponding solar abundance \citep{WilmsAM2000}. For the C6 observation, we extract the spectrum before the flux drops to almost zero. We fit the spectra with an absorbed power law \citep{EarnshawRH2016}. The hydrogen column density $N_{\rm{H}}$ is consistent with $\approx10^{21}$ cm$^{-2}$ except for that in the C6 observation. The photon indices are consistent with $\Gamma\approx1.5$ in all the data sets except for the dip phase in the C1 observation. \LD{The photon indices obtained in the non-dip states of all observations are fully consistent with those presented in \citet{EarnshawRH2016}.} The best-fit parameters are presented in the Table \ref{tab:chandra_spec}.

We calculate the absorbed model flux in the soft band (0.3 -- 2 keV) and hard band (2 -- 10 keV) using the \texttt{sample\_flux} command. Then we calculate the ratio between the flux at the hard band and the flux at the soft band.  The ratio of the non-dip state is stable at $\approx2.8$ and likely X-ray flux independent. All the ratios determined in the dip states remain consistent with those in the non-dip states although the uncertainties are clearly large. 

In the C7 observation, an excess of X-ray photons can still be seen in the position of ULX-7. The spectral parameters cannot be constrained if we set all parameters to be free. After freezing $N_{\rm{H}}$ to be $1.1\times10^{21}$~cm$^{-2}$, the photon index is constrained as $\Gamma=3\pm1$ with a statistic of $C^2/dof=13.8/20$. The X-ray luminosity can be estimated as $5^{+4}_{-2}\times10^{37}$~\lumcgs, which can be considered as the upper limit of the flux in the superorbital low state.

We noticed that spectra of PULXs can usually be fit with complex models containing more than one component \citep[see, e.g.,][]{KoliopanosVG2017}. Detailed spectral analysis with the \xmm\ data sets of X1 -- X4 suggests that the soft X-ray emission of ULX-7 can be empirically described by a low-temperature disk blockbody plus a high-temperature blackbody \citep{RodriguezCastilloIB2020}. We use this two-component model to fit the \chandra\ spectra of ULX-7 during non-dip states by linking $N_{\rm{H}}$ across all observations. The spectra can be fit with an inner disk temperature of $kT_{\rm{in}}\approx0.36$ keV and a hot blackbody with a temperature $kT_{\rm{BB}}\approx1.1$ keV across all the observations except for C6 which contains large uncertainties. Alternatively, the spectra can also be equally well described with other models, e.g., a cool blackbody plus a hot disk blackbody, or a cool disk blackbody plus a Comptonized corona. The implications of detailed spectroscopy, which is beyond the scope of this paper, have been discussed in \citet{RodriguezCastilloIB2020}. We attempted to fit the dip spectra with these two-component models, but did not observe significant differences in spectral parameters compared to non-dip spectra. 

\begin{table*}
    \caption{Best-fit parameters of \emph{Chandra} spectra of M51 ULX-7 with a single power law. }
    \label{tab:chandra_spec}
    \footnotesize
    \begin{tabular}{cccccccccc}
    \hline
    \hline
Epoch & State & \LD{Counts}  & $N_{\rm{H}}$  & $\Gamma$ & $L_{0.3\rm{-}10\rm{ keV}}$  & $F_{0.3\rm{-}2\rm{ keV}}$ & $L_{2\rm{-}10\rm{ keV}}$ & Ratio & $C^2/dof$\\
\hline
C1 & Non-dip & \LD{9047} & $1.3\pm0.2$ & $1.50\pm0.03$ & $76\pm3$ & $17.4\pm0.7$ & $50\pm3$ & $2.8\pm0.2$ & 508.2/498\\
C1 & Dip & \LD{1021} & $0.6_{-0.3}^{+0.5}$ & $1.32\pm0.07$ & $60_{-7}^{+8}$ & $13\pm2$ & $45\pm7$ & $3.3\pm0.5$ & 284.7/320\\
C2 & Non-dip & \LD{4208} & $1.1\pm0.2$ & $1.54\pm0.05$ & $63\pm3$ & $16\pm1$ & $40\pm3$ & $2.6\pm0.3$ & 402.2/442\\
C2 & Dip & \LD{3340} & $1.2\pm0.2$ & $1.48\pm0.05$ & $56\pm4$ & $13\pm1$ & $37\pm3$ & $2.9\pm0.3$ & 428.6/425\\
C3 & Non-dip & \LD{1827} & $0.8\pm0.3$ & $1.47\pm0.06$ & $53\pm5$ & $13\pm1$ & $36\pm4$ & $2.8\pm0.4$ & 375.3/377 \\
C4 & Non-dip & \LD{4629} & $1.6\pm0.3$ & $1.53\pm0.04$ & $50\pm3$ & $11\pm1$ & $33\pm3$ & $3.0\pm0.3$ & 457.4/449\\
C4 & Dip & \LD{1986} & $1.4\pm0.3$ & $1.68\pm0.06$ & $40\pm4$ & $10\pm1$ & $24\pm3$ & $2.4\pm0.4$ & 364.6/377\\
C5 & Non-dip & \LD{1745} & $1.1\pm0.3$ & $1.51\pm0.07$ & $31\pm3$ & $7.5\pm0.8$ & $20\pm3$ & $2.8\pm0.4$ & 323.7/371\\
C6 & Non-dip & \LD{479} & $<0.3$ & $1.4\pm0.1$ & $7.4_{-0.8}^{+0.9}$ & $2.0\pm0.2$ & $5.1\pm0.9$ & $2.5\pm0.5$ & 239.8/224\\
\hline
\tabularnewline
\end{tabular}
\\ {\footnotesize{The unit of  $N_{\rm{H}}$ is $10^{21}$ cm$^{-2}$; the unit of the unabsorbed luminosity $L_{0.3\rm{-}10\rm{ keV}}$ is $10^{38}$ erg s$^{-1}$; the unit of the absorbed flux is $10^{-13}$ erg s$^{-1}$ cm$^{-2}$. The column counts is the number of source counts, where the estimated background counts are subtracted. }}
\end{table*}
\begin{figure}
\centering
\includegraphics[width=0.9\linewidth]{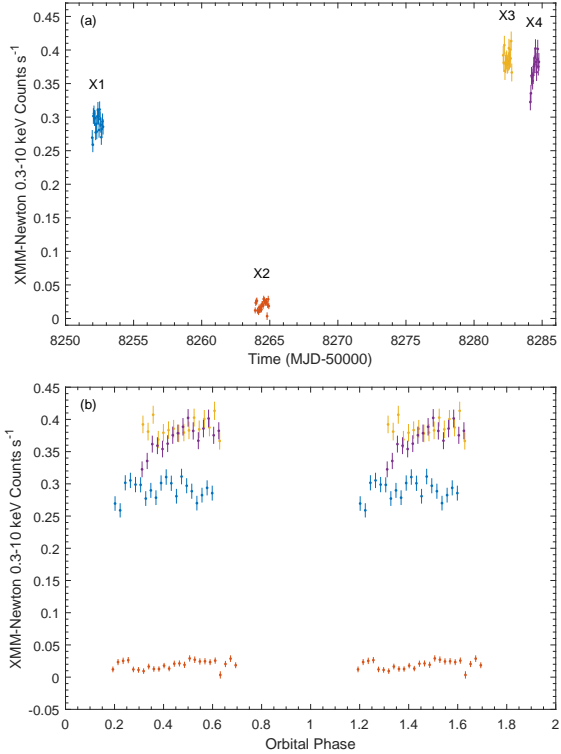} 
\caption{(a) \xmm\ 0.3–10 keV light curve of M51 ULX-7. The epochs of individual data sets are labeled. (b) Light curve folded with the orbital ephemeris \citep{RodriguezCastilloIB2020}. The orbital phases 0.7 -- 1.2 are not covered by \xmm\ observations. \label{fig:xmm_lc}}
\end{figure}

\subsection{\xmm\ Light Curves}
Figure \ref{fig:xmm_lc} (a) shows the XMM-Newton light curves using observations described in Table \ref{tab:observation_log} of ULX-7 in 0.3 -- 10 keV. X2 is likely observed in the superorbital low state, while the other three are taken in high or transition states. We observe no clear dip in the light curve. We then fold the light curve according to the orbital ephemeris and find that these four observations only cover orbital phases 0.2 -- 0.7 (see Figure \ref{fig:xmm_lc} (b)). This is expected because the orbital period of \xmm\ in 2018 is 47.86 hr (1.9942 days), which is very close to the orbital period of ULX-7 (1.9969 days). Given that only $\sim$40 hr are available for observation in each XMM-Newton orbit, it is impossible for the orbital phase of ULX-7 to be fully covered in a $\sim$40-day time span. Nevertheless, the \xmm\ observation suggests that the dips are unlikely to occur at orbital phases 0.2 -- 0.7, implying that the dips may occur near the phase of the superior conjunction of the pulsar. 

\section{Discussion}
The dips in Roche-lobe-filled X-ray binary systems are believed to be caused by an obscuring effect. The X-ray emission from the compact object and the inner accretion disk is obscured by the bulge on the rim of the accretion disk or the ring on the circularization radius \citep{NaikPA2011}. The absorption is strong in the soft X-ray band and a spectral hardening is expected. Currently known PULXs with short orbital and superorbital periods are likely driven by the Roche-lobe overflow although some wind-driven models are not entirely excluded \citep{RodriguezCastilloIB2020, TownsendC2020}. These Roche-lobe overflow high-mass X-ray binaries (HMXBs) could show orbital-phase-dependent dips that can be interpreted as the obscuring effect by the bulge similar to low-mass X-ray binaries \citep{NaikPA2011, Hu2013}.  In addition, the large size and the extended atmosphere of the companion star in a wind-fed HMXB make it possible to have an atmospheric eclipse that behaves like a dip and the X-rays from the NS are not fully obscured \citep{GrundstromBG2007, HuCN2017}. A high inclination angle is needed for both scenarios. However, the lack of a hardness change in ULX-7 during the dips is puzzling. This could be interpreted as a result of insufficient photon statistics, partial covering of the extended soft X-ray emission, or an extremely high optical depth of the absorber in the \chandra\ energy band. Given that the disk and the disk wind could contribute supersoft emission from a more extended region compared to the pulsed emission, a softening during the dip is possible if the supersoft emission is less obscured than the hard emission coming from the NS \citep{ZhouFH2019}. This partial covering of an extended soft emitter is also used to interpret the lack of hardness ratio change in  NGC 55 X-1 \citep{StobbartRW2004}, although the emission mechanism for the soft emission in ULX-7 is quite different \citep{StobbartRW2004}. Another possible scenario for the dip is the intrinsic variability of the mass accretion rate although this is not likely orbital dependent. 

%
\begin{figure}
\centering
\includegraphics[width=0.9\linewidth]{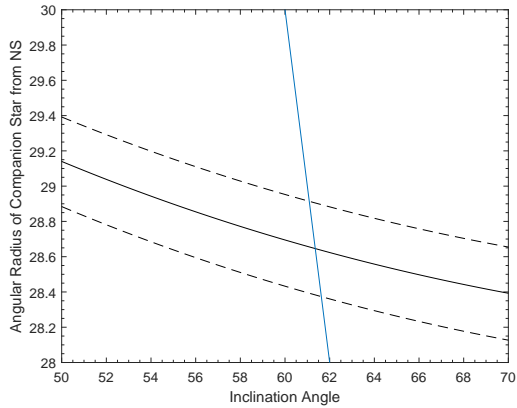} 
\caption{Black line: angular size $\phi$ of the companion star as a function of the inclination angle $i$ under the assumption that the companion star fills its Roche lobe. Dashed lines: 1-$\sigma$ intervals. Blue line: critical value for the Companion star to pass through the line of sight. \label{fig:inclination_angle}}
\end{figure}

If we further assume that the complement of the inclination angle ($i$) is not very far from the angular size of the companion star, the inclination angle could be constrained. The effective Roche-lobe size ($r_{\rm{L}}$) can be estimated as
\begin{equation}
    \frac{r_{\rm{L}}}{a}=\frac{0.49 q^{2/3}}{0.6q^{2/3}+\ln{(1+q^{1/3})}},
\end{equation}
where $q$ is the mass ratio and $a$ is the semi-major axis of the orbit of the compact object \citep{Eggleton1983}. Since the mass function and the semi-major axis of ULX-7 have been derived with pulsar timing as a function of $i$ \citep[see][]{RodriguezCastilloIB2020}, $r_{\rm{L}}$ as well as the angular size of the companion star ($\phi$) are functions of $i$ (see Figure \ref{fig:inclination_angle}). The blue line in Figure \ref{fig:inclination_angle} denotes $i=90^{\circ}-\phi$, which is the critical value of the occurrence of the eclipse. The intersection of $\phi(i)$ and this blue line suggest that the critical value of the inclination angle is $\sim62^{\circ}$, which is the upper limit of the inclination angle of ULX-7. Since we did not observe total eclipse, the inclination angle is likely slightly lower than this critical value (Figure \ref{fig:inclination_angle}).

A moderate geometric beamed funnel with a half-opening angle of $\sim60^{\circ}$ could explain the observed luminosity. This could also explain the presence of dips and the precession of the funnel could explain the superorbital modulation \citep{RodriguezCastilloIB2020}. Assuming that the beamed emission is symmetric to the disk plane, the $60^{\circ}$ half-opening angle of each side corresponds to a total solid angle of $2\pi$, which is half of the entire sphere. This enhances the observed flux by a factor of two compared to an isotropic emitter with the same accretion luminosity. The beaming factor can be derived as $b\equiv L_{\rm{acc}}/L_{\rm{iso}} \gtrsim1/2$, where $L_{\rm{acc}}$ is the total accretion luminosity and $L_{\rm{iso}}$ is the observed luminosity by assuming isotropic emission \citep{King2008}. This implies a peak column accretion luminosity of $L_{\rm{acc, max}}\approx4\times10^{39}$ erg s$^{-1}$. The maximum accretion luminosity can be approximated as
\begin{equation}\label{eq:bfield_boost}
    L_{\rm{acc,max}}\approx0.35\times\left( \frac{B}{10^{12}\rm{ G}} \right)^{3/4}\times10^{39}\rm{~erg~s}^{-1},
\end{equation}
where $10^{13}$ G $<B<10^{15}$ G if the high luminosity is powered by a strong magnetic field \citep{MushtukovST2015}. Therefore, a magnetic field of $B\gtrsim3\times10^{13}$~G is needed for ULX-7. We note that this equation is only an approximation and is based on a few assumptions, e.g., the accretion column height cannot be larger than the NS radius. Thus, equation \ref{eq:bfield_boost} can only be used as an order of magnitude estimation for the maximum magnetic field. Finally, the spectrum of ULX-7 can be described by two components \citep{RodriguezCastilloIB2020}, while the hard one likely represents the pulsed emission. Given that the pulsed emission only contributes $\sim$60~\% of the total luminosity, the lower limit of the magnetic field could be as low as $1.5\times10^{13}$ G. This remains strong enough to exclude a boost from a pure supercritical accretion model without a strong ($>10^{13}$~G) magnetic field. However, the funnel-shaped wind could be misaligned with respect to the normal factor of the disk plane. This makes if difficult to find a one-to-one mapping between the inclination angle and the beaming factor in PULXs. Further simulation and observational properties will be needed to constrain their relationship.

It remains possible that the stellar wind of the companion star of ULX-7 plays an important role, and the dip could be an absorption effect of wind clumps similar to those of Cyg X-1 and SMC X-2 \citep{FengC2002, LiHL2016}. This type of dip can happen at any orbital phase. If this is the case for ULX-7, the inclination angle cannot be well constrained. The orbital phase distribution of this kind of dip is usually less stable than those originating from the disk rim. Therefore, more observations and statistical analysis of dips in ULX-7 are needed to test their origin and constrain the viewing geometry of ULX-7.

\section{Summary}

Utilizing the archival \chandra\ data taken in 2012, we find that ULX-7 shows periodic dips in its decreasing trend of a superorbital cycle. Through epoch folding analysis and phase dispersion minimization, we find that the period of the dip is consistent with the orbital period. If the dip is caused by the absorption of the X-ray emission by the vertical structure on the stream-disk interaction region or the atmosphere of the companion star, the inclination angle of this system could be constrained as not very far from $\sim60^{\circ}$. This can be achieved if the magnetic field of the NSb is as high as $10^{13}$ G, which has been inferred from previous studies \citep{VasilopoulosLK2020}. An absorption from the clumpy wind may also be responsible for the origin of the dips, but they are expected to be less stable compared to those dips caused by the bulge on the disk. Moreover, the lack of hardening during the dip is puzzling with regard to their origin. More observations are needed to investigate the stability, orbital phase dependence, and the spectral behaviors of the dips. 

\section{Acknowledgements}
\LD{We thank the anonymous reviewer for valuable comments that improved this paper}. This research is in part based on the data obtained from the Chandra Data Archive and has made use of the software provided by the Chandra X-ray Center (CXC) in the application packages CIAO, ChIPS, and Sherpa. This research has used the observations obtained with XMM-Newton and the ESA science mission with instruments and contributions directly funded by the ESA member states and NASA. This research made use of \texttt{Astropy}, a community effort to develop a common core package for astronomy in Python. C.-P.H. acknowledges support from the the Ministry of Science and Technology in Taiwan through grant MOST 109-2112-M-018-009-MY3 and the support from Japan Society for the Promotion of Science (JSPS; ID: P18318).

\facilities{\emph{Chandra} (ACIS), \emph{XMM} (MOS, PN)}
\software{CIAO \citep{FruscioneMA2006}, \xmm\ SAS \citep{GabrielDF2004}, Astropy\citep{astropy2013, astropy2018_1, astropy2018_2}}


\end{document}